\documentclass[
reprint,
amsmath,
amssymb,
aps,
prapplied,
]{revtex4-2}

\usepackage{hyperref}
\usepackage[capitalize]{cleveref}

\usepackage{graphicx}
\usepackage{dcolumn}
\usepackage{bm}
\usepackage{physics}
\usepackage{nameref}
\usepackage{upgreek}
\usepackage{amsmath}
\usepackage{siunitx}
\usepackage{braket}
\usepackage{upgreek}

\begin{document}

\title{Optimization of High-Fidelity Single-Qubit Gates for Fluxoniums Using Single-Flux Quantum Control}

\author{%
	Maxime~Lapointe-Major,\textsuperscript{1,2,3} Boyan~Torosov,\textsuperscript{1} Bohdan~Kulchytskyy,\textsuperscript{1} and Pooya~Ronagh\textsuperscript{1,4,5,6,}
}
\thanks{Corresponding author: \href{mailto:pooya.ronagh@1qbit.com}{pooya.ronagh@1qbit.com}}

\affiliation{\textsuperscript{\emph{1}}1QB Information Technologies (1QBit), Vancouver, BC, Canada\\
\textsuperscript{\emph{2}}D\'{e}partement de g\'{e}nie \'{e}lectrique et de g\'{e}nie informatique, Universit\'{e} de Sherbrooke, Sherbrooke, QC, Canada\\
\textsuperscript{\emph{3}}Institut Quantique, Universit\'{e} de Sherbrooke, Sherbrooke, QC, Canada\\
\textsuperscript{\emph{4}}Institute for Quantum Computing, University of Waterloo, Waterloo, ON, Canada\\
\textsuperscript{\emph{5}}Department of Physics \& Astronomy, University of Waterloo, Waterloo, ON, Canada\\
\textsuperscript{\emph{6}}Perimeter Institute for Theoretical Physics, Waterloo, ON, Canada
}

\date{\today}

\begin{abstract}
We present a gradient-based method to construct memory-efficient, high-fidelity, single-qubit gates for fluxonium qubits. These gates are constructed using a sequence of single-flux quantum (SFQ) pulses that are sent to the qubit through either capacitive or inductive coupling. The schedule of SFQ pulses is constructed with an on-ramp and an off-ramp applied prior to and after a pulse train, where the pulses are spaced at intervals equal to the qubit period. We reduce the optimization problem to the scheduling of a fixed number of SFQ pulses in the on-ramp and solve it by relaxing the discretization constraint of the SFQ clock as an intermediate step, allowing the use of the Broyden--Fletcher--Goldfarb--Shanno optimizer. Using this approach, gate fidelities of 99.99\% can be achieved for inductive coupling and 99.9\% for capacitive coupling, with leakage being the main source of coherent errors for both approaches.
\end{abstract}

\maketitle

\section{Introduction\label{sec:intro}}
Superconducting qubits are a promising candidate towards the physical implementation of a fault-tolerant quantum computer~\cite{blais2021circuit, kjaergaard2020superconducting}. Their scalable circuit designs offer a wide range of qubits, such as transmons~\cite{koch2007charge, arute2019quantum}, flux qubits~\cite{harris2010experimental}, and fluxoniums~\cite{manucharyan2009fluxonium, nguyen2022blueprint, nguyen2019high, somoroff2023millisecond, earnest2018realization, rower2024suppressing, ding2023high}, each of which have their respective advantages. Traditionally, the control of these qubits has been realized by delivering classical microwave pulses generated at room temperature onto a qubit through wires. However, a number of challenges are foreseen when scaling up the number of physical qubits onto quantum processors, such as heat dissipation~\cite{krinner2019engineering}, connectivity logistics, and physical space for the wires and the control and readout devices.

An alternative approach to using classical microwave in qubit control is to use single-flux quantum (SFQ) pulses~\cite{lin1995timing, mcdermott2014accurate, mcdermott2018quantum}. In SFQ control schemes, control signals are generated on a cryogenic chip placed in a dilution refrigerator with the quantum processor to be controlled. These control signals are discrete, short pulses with a precise integral equal to a single-superconducting-flux quantum $\Phi = h/2e$, which are delivered to the qubit using a classical control clock co-integrated on the control chip via either inductive or capacitive coupling. This type of scalable solution to qubit control has been used on transmons to implement single- and two-qubit gates with fidelities comparable to those of traditional control schemes~\cite{mcdermott2014accurate, torosov2025optimization, shillito2025compact, leonard2019digital, li2019hardware, howe2022digital, liu2023single, liebermann2016optimal, mcdermott2018quantum, dalgaard2020global, wang2023single, jokar2021practical}.

In our work, we extend the use of this new and promising technology to another type of superconducting qubits, namely, the fluxonium, which has emerged as a promising qubit due to its large anharmonicity and long coherence times~\cite{nguyen2022blueprint, nguyen2019high, somoroff2023millisecond}. A number of approaches can be used for the implementation of an SFQ pulse schedule. The most na\"ive approach is to apply a pulse train of evenly spaced pulses onto a qubit, which within the rotating wave approximation is equivalent to a square pulse applied using traditional microwave control~\cite{leonard2019digital, shillito2025compact}. This na\"ive approach has the benefit of being memory-efficient for the encoding of pulse schedules, but yields poor gate fidelities. Another approach found in the literature involves optimizing pulse schedules using genetic or trust-region algorithms~\cite{liebermann2016optimal, vogt2022binary}. This approach usually achieves high fidelities, but requires an inefficient bit-by-bit encoding of the resulting pulse schedule. In this work, we propose a hybrid approach similar to Ref.~\cite{shillito2025compact}, in which the SFQ pulse schedule is constructed using a pulse train and both an on-ramp and an off-ramp applied prior to and after the pulse train, respectively. By imposing the appropriate restrictions onto the ramp construction, high fidelities can still be achieved with minimal compromise in terms of memory requirements. In particular, we fix the off-ramp to mirror the on-ramp with respect to time, and optimize the pulse schedule for a fixed ramp length and a fixed number of pulses within the ramp. This optimization is performed first by relaxing the time discretization constraint from the SFQ control clock, which allows the use of the Broyden--Fletcher--Goldfarb--Shanno (BFGS) optimizer~\cite{fletcher2000practical} to find a high-fidelity pulse schedule, followed by the reactivation of the constraint to snap the SFQ pulses into alignment with ticks of the SFQ clock. We apply our ramp method to both inductive and capacitive couplings between the SFQ control chip and the fluxonium.

This paper is organized as follows. In~\Cref{sec:model}, a model for a fluxonium qubit and the framework of SFQ control are introduced. \Cref{sec:optimization} details how the ramp optimization problem is broken down and then solved. \Cref{sec:results} presents the fidelities obtained for the proposed control scheme, along with a detailed error budget for our optimized pulse schedules. \Cref{sec:encoding} describes how the pulse schedule can efficiently be programmed on a control chip.

\section{Model\label{sec:model} }
The model considered for the fluxonium qubit consists of a capacitance, an inductance, and a Josephson junction in parallel~\cite{nguyen2022blueprint} (see~\cref{fig:circuitschematic}). The Hamiltonian of the qubit can be written as
\begin{equation}
    \hat{H}/h = 4E_\mathrm{C}\hat{n}^2 - E_\mathrm{J} \mathrm{cos} ( \hat{\phi} + \phi_{\mathrm{ext}}) + \frac{1}{2}E_\mathrm{L} \hat{\phi}^2 \label{eq:HFock}, 
\end{equation}
where $\phi_\mathrm{ext}$ is the external magnetic flux applied to the superconducting loop, $E_\mathrm{C}$ is the charging energy, $E_\mathrm{J}$ is the Josephson energy, $E_\mathrm{L}$ is the inductive energy, and $\hat{n}$ and $\hat{\phi}$ are the charge and phase operators, respectively. Using the ladder operators $\hat{b}^\dagger$ and $\hat{b}$, the charge and phase operators can be written as
\begin{equation}
\hat{\phi} = \frac{1}{\sqrt{2}} \left( \frac{8E_\mathrm{C}}{E_\mathrm{L}}\right)^{1/4}(\hat{b}^\dagger + \hat{b}) \label{eq:phase},
\end{equation}
\begin{equation}
\hat{n} = \frac{i}{\sqrt{2}}\left( \frac{E_\mathrm{L}}{8E_\mathrm{C}} \right)^{1/4}(\hat{b}^\dagger - \hat{b}) \label{eq:charge}.
\end{equation}
A fluxonium is obtained by choosing the parameter configuration of $\{E_\mathrm{J}, E_\mathrm{C}, E_\mathrm{L}\} = \{4, 1, 1\}$ GHz and $\phi_\mathrm{ext}$ of half a quantum of flux, which leads to the diagonal Hamiltonian $\hat{H}_0 / \hbar = \sum_j \omega_j \ket{j}\bra{j}$. Here, the qubit frequency is $\omega_{01}/2\pi=0.58$ GHz and the energy gap to the first leakage state is $\omega_{12}/2\pi=3.39$ GHz. The energy levels used in our numerical simulations are computed by diagonalizing the Hamiltonian of~\cref{eq:HFock} using 30 states in the Fock basis and projecting the resulting energy levels onto the six lowest-energy states.

The coherence times used for open-system simulations, taken from Ref.~\cite{somoroff2023millisecond}, are given by $T_\mathrm{1}=\SI{1.2}{\milli\second}$ and \mbox{$T_\mathrm{2} =\SI{800}{\micro\second}$.} Thermal excitations have been omitted.

\subsection{Qubit Control Using SFQ Pulses\label{subsec:sfq} }

\begin{figure}[h]
        \includegraphics[width=\linewidth]{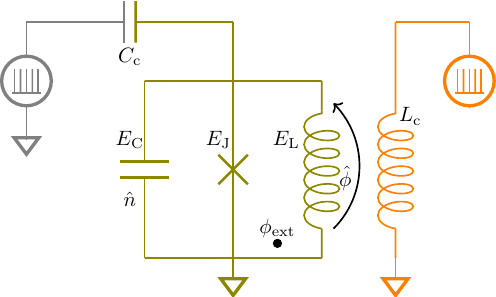}
        \caption{Circuit diagram of a fluxonium (shown in green) either inductively coupled to the SFQ generator (orange) or capacitively coupled to the SFQ generator (grey). }
        \label{fig:circuitschematic}
\end{figure}

The control scheme considered is a sequence of discrete, fixed-amplitude, fast pulses applied to the fluxonium on a control clock, referred to as the SFQ clock. \Cref{fig:circuitschematic} shows a circuit diagram of an SFQ-driven fluxonium (shown in green) through an inductive coupling (orange) or a capacitive coupling (grey). 
In practice, SFQ pulses have a finite width ($\approx \SI{1}{\pico\second}$), but this duration is negligible compared to the other timescales we consider. This leads to the accurate approximation of an SFQ pulse by a Dirac $\delta$ function~\cite{mcdermott2014accurate}. The corresponding driving Hamiltonians for inductive and capacitive couplings are
\begin{equation}
\hat{H}_\mathrm{SFQ}^\mathrm{L} = -\frac{L_\mathrm{c}}{L'}I(t)\hat{\phi},
\end{equation}
\begin{equation}
\hat{H}_\mathrm{SFQ}^\mathrm{C} = -\frac{C_\mathrm{c}}{C'}V(t)\hat{n},
\end{equation}
respectively, where $L_\mathrm{c}$ ($C_\mathrm{c}$) is the coupling inductance (capacitance) and $L'$ ($C'$) is the self-inductance (self-capacitance) of the fluxonium. By casting this Hamiltonian in the Schr\"{o}dinger equation and using $\int^\infty_{-\infty}I(t')dt'=2e$, the charge carried by a Cooper pair, the unitary resulting from a kick can then be written as
\begin{equation}
\hat{U}_\mathrm{kick}^\mathrm{L} (\theta_\mathrm{L}) = \mathrm{exp}\left( \frac{i\theta_\mathrm{L}}{2|\phi_{01}|} \hat{\phi}\right), 
\end{equation}
\begin{equation}
\hat{U}_\mathrm{kick}^\mathrm{C} (\theta_\mathrm{L}) = \mathrm{exp}\left( \frac{i\theta_\mathrm{C}}{2|n_{01}|} \hat{n}\right), 
\end{equation}
with $|\phi_{01}| = \bra{0}\hat{\phi} \ket{1}$, $|n_{01}| = \bra{0}\hat{n} \ket{1}$, and $\theta_\mathrm{L} \propto L_\mathrm{c}/\sqrt{L'}$ ($\theta_\mathrm{C} \propto C_\mathrm{c}/\sqrt{C'} $) is the kick angle in radians. The ``01'' matrix element is a rescaling element that ensures the kick angle $\theta_\mathrm{L}$ ($\theta_\mathrm{C}$) is in units of radians per kick.

The evolution of the qubit following a pulse train can then be described as a series of operations consisting of SFQ kicks and free evolutions given by
\begin{equation}
\hat{U} = \prod_{j=0}^{N} \mathrm{exp}(-i\hat{H}_0 t_j)\hat{U}_\mathrm{kick},
\end{equation}
where $N$ is the number of kicks and $t_j$ is the time between the $j$-th kick and the $(j+1)$-th kick.

By choosing an idling time equal to the qubit period $T = 2\pi / \omega_{01}$ and a number of kicks $N = \theta_\mathrm{targ}/\theta_{(\mathrm{L},\mathrm{C})}$, arbitrary rotations $\hat{R}_{(-X)}(\theta_\mathrm{targ})$ and $\hat{R}_{(-Y)}(\theta_\mathrm{targ})$ can be approximated for inductive and capacitive SFQ coupling, respectively.  

\begin{figure}[h]
        \includegraphics[width=\linewidth]{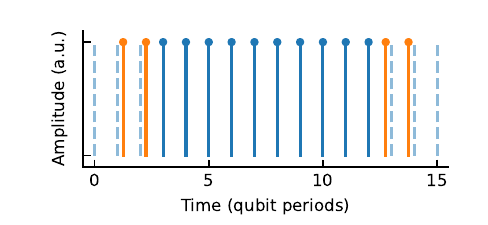}
        \caption{Schematic representation of an SFQ pulse schedule with a ramp length of three qubit periods, two pulses within each ramp, and a pulse train comprising 10 pulses.}
        \label{fig:scheduleschematic}
\end{figure}

We choose to implement arbitrary rotations using a train pulse of $N_\mathrm{train}$ kicks, each separated by the qubit period, and an on-ramp and an off-ramp applied before and after the pulse train, respectively. The frequency of the pulse train provides energy selectiveness for the pulse schedule. This is usually achieved by multiplying the control signal by a cosine with a carrier frequency equal to the frequency of the qubit in traditional microwave control schemes. In addition, the on-ramp and off-ramp allow the implementation of a DRAG (derivative removal by adiabatic gate)~\cite{motzoi2009simple} equivalent in the pulse schedule~\cite{shillito2025compact}, which can minimize leakage and compensate for the Stark shift~\cite{nguyen2022blueprint, xiong2022arbitrary} on the fluxonium resulting from the application of the pulse schedule. \Cref{fig:scheduleschematic} shows a schematic representation of an SFQ pulse schedule.

\section{Optimization Method\label{sec:optimization} }
We reduce the problem of finding the pulse schedule to a ramp optimization followed by a pulse train length optimization. The off-ramp applied after the pulse train does not require optimization as we choose to fix it to mirror the on-ramp, decreasing the size of the optimization problem. Despite this simplification, the optimization problem remains fairly large because of the high clock frequencies used in state-of-the-art experiments ($>$$\SI{100}{\giga\hertz}$) and the low fluxonium qubit frequency relative to that of the SFQ clock. For an SFQ clock at 128$\times$ the qubit's frequency ($\approx \SI{75}{\giga\hertz}$) and the maximum ramp length of five qubit periods considered in this work, the optimization problem has 640 binary variables. However, given the kick angle that will be chosen prior to optimization, we wish to limit the number of pulses within its ramp for efficient sequence encoding and to maintain the energy selectiveness the pulse train brings to the overall pulse schedule. Therefore, only solutions with up to six SFQ kicks within a ramp are considered for a ramp of any length.

\subsection{Ramp Optimization\label{subsec:ramp} }
Our ramp optimization begins by reducing the problem's size by setting both a fixed ramp length and a fixed number of pulses within the ramp. Next, the SFQ clock's time discretization constraint is relaxed, which allows the arrival time of the SFQ pulses to be treated as continuous variables. This allows continuous optimization tools to find a high-fidelity SFQ pulse schedule. Here, we choose the BFGS algorithm, which is a quasi-Newton method that has shown acceptable performance for similar problems~\cite{torosov2025optimization}.

For a given choice of kick and target angles, a ramp optimization is performed for each combination of the number of SFQ pulses $N=1,2,\ldots ,6$ and the total duration of ramp lengths $t(R)=RT$, where $R=1,2,\ldots ,5$ and $T=2\pi/\omega_{01}$ is the qubit period. Following each choice of $N$ and $R$, the ramp is optimized, and the best result among all combinations of the two is returned.

For any fixed ramp, the number of pulses in the pulse train $N_\mathrm{train}$ is optimized using an exhaustive search. The associated cost function is the process infidelity $1-F_\mathrm{pro}$ of the pulse schedule concatenating the pulse train and the mirrored copies of the ramp. Here,
\begin{equation}
F_\mathrm{pro} = \frac{1}{4} \left|\mathrm{Tr}[\hat{U}^\dagger_\mathrm{targ} \hat{U}_Q]\right|^2, \label{eq:ProcessFid}
\end{equation}
where $\hat{U}_\mathrm{Q}$ is the projection of the full unitary $\hat{U}$ onto the computation subspace and $\hat{U}_\mathrm{targ}$ is the target unitary.

Our optimization landscape as a function of SFQ pulse timing within a ramp is highly non-convex, and contains many local minima due to the discretization of the pulse train length, but also peaks with a spacing of $T/6$ due to an unknown physical source. In addition, in the case of capacitive coupling, global minima are generally found for SFQ pulse timings near $t~(\mathrm{mod}~T) = T/4$, which is equivalent to sending a DRAG component in the control signal in traditional microwave control~\cite{shillito2025compact}. In the case of inductive coupling, global minima are found for solutions between $t~(\mathrm{mod}~T) = 0$ and $t~(\mathrm{mod}~T) = T/4$; having a pulse timing in this range is equivalent to sending a small but not maximized DRAG component in the control signal.

Given these observations, our method uses many BFGS trials. For capacitive coupling, the ensemble of initial conditions consists of all of the non-unique combinations of pulse timings at (1) $rT+T/4$, $r=0, 1, \ldots ,R-1$; and (2) the results from the optimization for $N-1$ and $R-1$, with an additional pulse placed at $(R-1)T$. The latter ensures that, in the worst case, the same result is found as in the previous optimization, but with a pulse train having two fewer pulses, since those pulses are moved into the now-longer ramps. For inductive coupling, the ensemble of initial conditions consists of all of the non-unique combinations of pulse timings at (1) $rT + T/20$, $r=0, 1, \ldots ,R-1$; (2) $rT + T/4$, $r=0, 1, \ldots ,R-1$; (3) $rT + 3T/4$, $r=0,1,\ldots ,R-1$; and (4) results of the optimization for $N-1$ and $R-1$, with an additional pulse placed at $(R-1)T$. Last, for both couplings, if the number of pulses considered is greater than the number of non-unique combinations thus far considered, a set of $N$ evenly spaced pulses between 0 and $R$ are added to the list of pulse timings from which all of the non-unique combinations are then computed again in order to find the new ensemble of initial conditions. 

After the best solution has been found within the considered ensemble of initial conditions, a heuristic for exploring neighbouring solutions is used to ensure that no better solution can be found nearby. This heuristic is implemented to navigate the peaks with a spacing of $T/6$. For each of the $N$ pulses, a jump of $\pm T/6$ is applied, yielding three initial conditions for each pulse. The resulting $3^N$ combinations of new initial conditions is then passed to the BFGS optimizer. This heuristic is repeated iteratively until convergence has been achieved.

After the optimization process using a continuous-time variable for the arrival time of the SFQ pulses has been completed, the SFQ clock's time discretization constraint is applied by forcing the SFQ pulses to snap onto (i.e., align with) ticks of the SFQ clock. To do so, pulses are moved onto their neighbour SFQ clock ticks such that the gap to the solution of the continuous-variable problem is minimized. However, if there are more than two pulses between two neighbouring ticks of the SFQ clock in the continuous-time solution, the ramp is discarded. This ensures that schedules with degenerate kicks, which can occur more frequently in the case of small ramps consisting of a large number of pulses (e.g., $R=1$, $N=6$), are discarded to avoid a large jump in one or more of the pulses' arrival time, leading to a large jump in the resulting gate infidelity.

\section{Results\label{sec:results} }
\begin{figure}[h]
        \includegraphics[width=\linewidth]{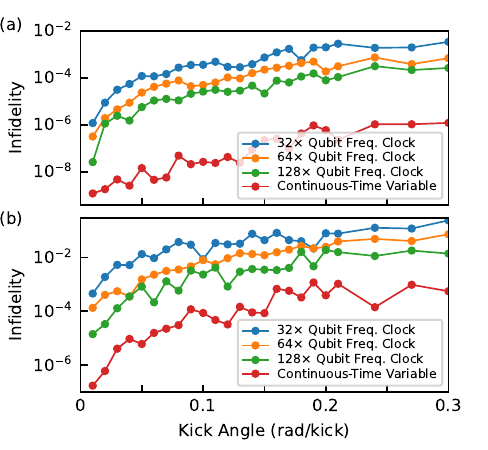}
        \caption{Best infidelity obtained as a function of the kick angle with a relaxed SFQ clock constraint (i.e., optimized with respect to continuous variables for SFQ kick arrival times) and for different SFQ clock frequencies in the case of (a) inductive coupling and (b) capacitive coupling. For each kick angle, the pulse schedule is optimized for ramps of $R=1,2, \ldots, 5$ qubit periods in length, for $N=1, 2, \ldots, 6$ pulses, and for a target rotation angle of $\theta_\mathrm{targ}=\pi$.}
        \label{fig:kickanglesweep}
\end{figure}

The first physical parameter for which SFQ pulse schedules are optimized is the kick angle $\theta$. For these optimizations, the target rotation angle $\theta_\mathrm{targ}$ is a $\pi$-pulse, and the maximum number of pulses in the ramp is $N=6$. \Cref{fig:kickanglesweep} shows the best results obtained as a function of the kick angle for both the inductive and the capacitive couplings. Given the coherence times of fluxoniums and the gate times resulting from SFQ control, incoherent errors are expected to be on the order of $10^{-4}$ and, therefore, a kick angle where the infidelity from the coherent errors is of the same order of magnitude is chosen. This leads us to use the values $\theta_\mathrm{L} = 0.15$ and $\theta_\mathrm{C} = 0.03$, as infidelities in the range of the desired $10^{-4}$ threshold are attainable for a 128$\times$ SFQ clock.

\begin{figure}[h]
        \includegraphics[width=\linewidth]{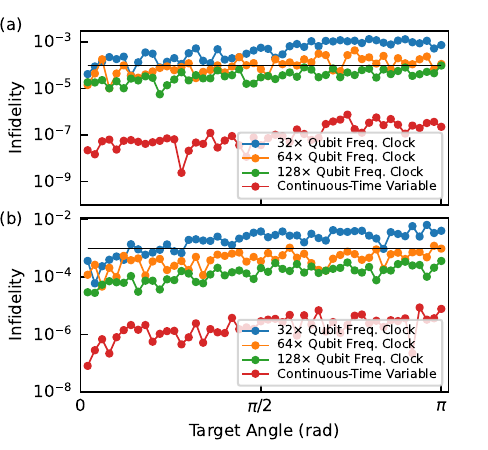}
        \caption{Best infidelity obtained as a function of the target rotation angle with a relaxed SFQ clock constraint (i.e., optimized with respect to continuous variables for SFQ kick arrival times) and for different SFQ clock frequencies in the case of (a) inductive coupling with a kick angle of $\theta_\mathrm{L} = 0.15$ radians per kick and (b) capacitive coupling with a kick angle of $\theta_\mathrm{C} = 0.03$ radians per kick. For each target angle, the SFQ ramp is optimized for ramps of $R=1, 2, \ldots, 5$ qubit periods in length, for $N=1, 2, \ldots, 6$ pulses.}
        \label{fig:rotanglesweep}
\end{figure}

\Cref{fig:rotanglesweep} shows the best infidelity obtained as a function of the target rotation angle $\theta_\mathrm{targ}$ for the selected kick angles. When considering a pulse schedule with a relaxed SFQ clock constraint (i.e., a continuous-time variable), infidelities of about $10^{-7}$ can be reached in the case of inductive coupling, and $10^{-6}$ in the case of capacitive coupling. Both control schemes show an increase in infidelity as a function of the target rotation angle $\theta_\mathrm{targ}$. As the target rotation angle increases, the number of pulses in the pulse train also increases, which causes some of the error rates to cumulatively increase. In addition, irregular jumps occur throughout the plot. These jumps are explained by the fact that the optimization process can be stuck in local minima. Furthermore, snapping the SFQ pulses onto ticks of the SFQ clock causes a large jump in infidelity. For a 128$\times$ clock frequency, infidelities of about $10^{-4}$ can be reached in the case of inductive coupling and $10^{-3}$ in the case of capacitive coupling. Also plotted are the resulting infidelities for a 64$\times$ and a 32$\times$ clock frequency, demonstrating that there is a gain in infidelity as the clock frequency is increased.
\subsection{Error Budget\label{sec:budget} }

\begin{figure*}[ht]
        \includegraphics[width=\linewidth]{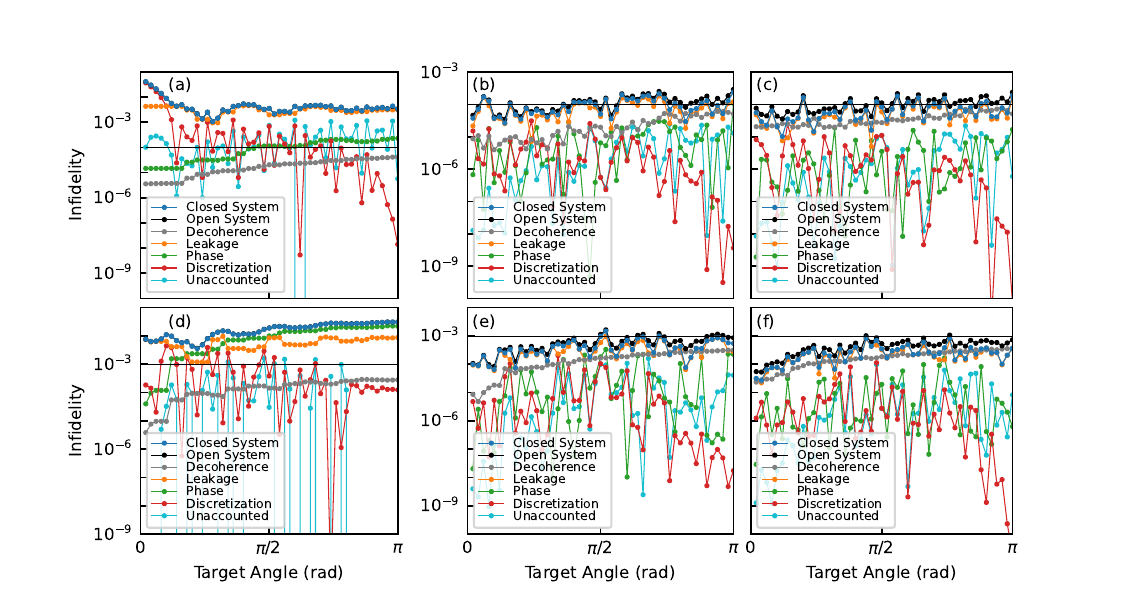}
        \caption{Error budget as a function of the target rotation angle $\theta_\mathrm{targ}$ for an SFQ clock at 128$\times$ the qubit frequency. (a)--(c) Results in the case of inductive coupling with a kick angle of $\theta_\mathrm{L} = 0.15$ radians per kick and (d)--(f) in the case of capacitive coupling with a kick angle of $\theta_\mathrm{C} = 0.03$ radians per kick. The left column ((a) and (d)) shows the error budget for a pulse train without ramps, the middle column ((b) and (e)) for a fixed ramp length of $R=1$, and the right column ((c) and (f)) for a fixed ramp length of $R=5$. The open-system infidelity (show in black) for the implementation without ramps cannot be seen because it falls under the closed-system infidelity (blue).}
        \label{fig:errorbudget}
\end{figure*}

\Cref{fig:errorbudget} shows the error budgets for pulse trains without any ramps and for gates constructed using fixed ramp lengths of $R=1$ and $R=5$, for both inductive and capacitive couplings. The closed system infidelity, computed from optimized pulse schedules using~\cref{eq:ProcessFid} is shown in blue. The leakage, computed for a closed system, is given by
\begin{equation}
L_\mathrm{closed} = 1 - \frac{1}{2}\mathrm{Tr}(\hat{U}_Q\hat{U}_Q^\dagger)\label{eq:leakage},
\end{equation}
(shown in orange). Other sources of coherent errors considered are the phase error
\begin{equation}
\mathcal{E}_\mathrm{phase} = \left|c_Z\right|^2,
\end{equation}
(shown in green), and the discretization error, which results from building the continuous rotation on the Bloch sphere out of discrete SFQ kicks, leading to a mismatch between the realized $X$ ($Y$) rotation component in the Pauli basis and the target value (shown in red).  For inductive and capacitive couplings, it takes the form
\begin{equation}
\mathcal{E}^\mathrm{L}_\mathrm{disc} = \left|c_X - \mathrm{sin}(\theta_\mathrm{targ}/2)\right|^2,
\end{equation}
\begin{equation}
\mathcal{E}^\mathrm{C}_\mathrm{disc} = \left|c_Y - \mathrm{sin}(\theta_\mathrm{targ}/2)\right|^2,
\end{equation}
respectively. The coefficients $c_i$ are obtained by decomposing the projected operator $\hat{U}_Q$ into Pauli matrices according to 
\begin{equation}
\hat{U}_Q = (1-\delta)(c_I \hat{I} + c_X \hat{X} + c_Y \hat{Y} + c_Z \hat{Z}),
\end{equation}
with $\sum_i |c_i|^2 = 1$ and where $L_1 = 2\delta-\delta^2$ is an alternative parametrization of leakage to that used in~\cref{eq:leakage}. The unaccounted-for errors are computed as \mbox{$\mathcal{E}_\mathrm{unacc} = (1-F_\mathrm{pro}) - (L_\mathrm{closed} + \mathcal{E}_\mathrm{phase} + \mathcal{E}_\mathrm{disc})$,} and are shown in cyan.

For the pulse trains without any ramps, the infidelity is almost two orders of magnitude greater than for unitaries constructed using ramps. For inductive coupling (see~\cref{fig:errorbudget}(a)), leakage is the largest contributor to the infidelity, and the other sources of coherent errors are an order of magnitude smaller than the leakage, except in the small target rotation angle $\theta_\mathrm{targ}$ regime, where discretization dominates. However, for capacitive coupling (\cref{fig:errorbudget}(d)), the phase error is considerably greater. This phase error is caused by the Stark shift resulting from the application of the pulse train, which detunes the qubit frequency while the control signal is being applied onto the qubit. When ramps are added to the pulse schedule (\cref{fig:errorbudget}(b)--(c), (e)--(f)), a gain of almost two orders of magnitude in the infidelity is observed, along with a decrease in all of the sources of error. These improvements demonstrate that our proposed approach is quite efficient at compensating for the Stark shift and mitigating the leakage for both coupling schemes, independent of the chosen ramp length. Given the observed performance of the ramps on phase error, we also expect our ramp method to be effective on fluxoniums whose frequencies have become detuned from the frequency of the SFQ clock.

While including ramps in a pulse schedule is  efficient at reducing leakage by more than an order of magnitude, this error rate remains the principal source of coherent error. This result is due to the fact that leakage comes from two physical sources, one of which cannot be addressed by the use of ramps alone. With respect to the first source, the spectral density of the pulse schedule cannot be efficiently minimized at the leakage frequency $\omega_{12}=\SI{3.4}{\giga\hertz}$, which is much larger than the qubit frequency $\omega_{01}=\SI{580}{\mega\hertz}$. The spectral density of the pulse schedule supports a broad range of frequencies around the target frequency and has peaks at the harmonics of the pulse train frequency, resulting in non-negligible modes at the leakage frequency. With respect to the second physical source of leakage, the shape of the phase $\hat{\phi}$ and charge $\hat{n}$ operators in the energy eigenbasis is considerably different from the ladder operator basis construction of $(b^\dagger \pm b)$ from~\cref{eq:phase} and~\cref{eq:charge} in two ways. First, the amplitude of the transition matrix elements $\bra{0}\hat{\phi}\ket{1}$ and $\bra{0}\hat{n}\ket{1}$ are strongly attenuated, which leads to a considerable deviation from the expected $\sqrt{N}$ scaling of the off-diagonal terms between the qubit subspace and leakage states~\cite{krantz2019quantum}. In addition, both the charge and phase operators have non-negligible non-tridiagonal matrix elements~\cite{krantz2019quantum}. In particular, we note that $\hat{n}_{03} \approx 2 \hat{n}_{01}$, meaning that the ground state is more strongly coupled to the $\ket{3}$ state than the $\ket{1}$ state, and that the energy mismatch between $\omega_{01}$ and $\omega_{12}$ in the control signal is not sufficient to fully mitigate leakage.

Leakage is generally considered one of the most severe types of error, as it can propagate to neighbouring qubits via two-qubit interactions and produce correlated errors~\cite{miao2023overcoming}. However, recent advances in the field of quantum error correction (QEC) indicate that leakage might not be as great a hurdle as initially thought, as protocols for leakage removal can be implemented between QEC cycles~\cite{miao2023overcoming, mcewen2021removing}. Should it still be an issue, leakage could be further minimized in our proposed SFQ control scheme by further reducing the kick angle, which would come at the cost of longer pulse schedules, and therefore a larger contribution to the infidelity by incoherent errors. The issue could also be addressed by further minimizing the frequency content of a pulse schedule at the leakage frequency, which could be achieved by using a combination of longer ramps and a greater number of pulses within the ramps, but would come at the cost of a larger memory overhead in encoding the resulting pulse schedules.

The final source of error considered in the error budget is the open-system fidelity
\begin{equation}
F_\mathrm{open} = \frac{1}{4}\mathrm{Tr}\left(S_\mathrm{Q}^\dagger (P_2 \otimes P_2 )S_\mathcal{E} \right),
\end{equation}
where $S_\mathrm{Q}$ is the superoperator representation of the target unitary projected onto the computational subspace, $S_\mathcal{E}$ is the superoperator representation of the computed map, and the term $(P_2 \otimes P_2 )$ projects the superoperator onto the computational subspace. The superoperator $S_\mathrm{Q}$ is computed by replacing the unitary for free evolutions between SFQ pulses by a propagator that is computed using the QuTip library~\cite{johansson2012qutip} and taking into account loss of energy at a rate of $\Gamma_\mathrm{1} = 1/T_\mathrm{1}$ and dephasing at a rate of $\Gamma_\upphi = 1/T_\mathrm{2} - 1/2T_\mathrm{1}$. The resulting open-system infidelity $1-F_\mathrm{open}$ is shown in black in~\cref{fig:errorbudget}, from which the contribution of incoherent errors (grey) is computed as $F_\mathrm{open} - F_\mathrm{closed}$. We note that the open-system infidelity for the implementation without ramps (subpanels (a) and (d)) cannot be seen because it falls under the closed-system infidelity (blue). Incoherent errors are similar in magnitude to the sum of coherent errors (blue), demonstrating an appropriate choice of kick angle for both inductive and capacitive couplings. 

\section{Encoding the Pulse Sequence\label{sec:encoding} }

One of the key reasons to turn towards SFQ control schemes for qubit control is the scalability this technology provides. As such, the pulse sequence needs to be encoded efficiently to minimize memory and communication overheads. For the proposed pulse schedule construction and the maximum  SFQ clock frequency of 128$\times$ considered in this work, the full on-ramp schedule can be encoded in $N_\mathrm{max} \times \mathrm{log}_2(128R + 1)$ bits, where the ``$+~1$'' represents the possibility of the ramp having fewer than the maximum allowed number of pulses ($N < N_\mathrm{max}$). Because the pulse train has a predefined schedule, it requires the encoding of only the total pulse count within the pulse train, which can be achieved using $\mathrm{log}_2(N_\mathrm{train})$ bits. The ramp length itself can be encoded using $\mathrm{log}_2(R_\mathrm{max})$ bits. An entire pulse schedule can therefore be encoded using $N_\mathrm{max}\times \mathrm{log}_2(128R+1) + \mathrm{log}_2(N_\mathrm{train}) +\mathrm{log}_2(R_\mathrm{max})$ bits.  

For the maximum $N_\mathrm{max}=6$ pulses within the ramp and the maximum ramp length $R_\mathrm{max}=5$ qubit periods considered in this work, the SFQ pulse schedule can be encoded using $6\times \mathrm{log}_2(640+1) < 56$ bits and the ramp length using $\mathrm{log}_2(5) < 3$ bits. In the case of inductive coupling, the pulse train is approximately 20 pulses long for a $\pi$ pulse, while for capacitive coupling, the pulse train is about 100 pulses long. Therefore, 5 bits are sufficient to encode the number of pulses within the pulse train for inductive coupling while 7 bits are sufficient for capacitive coupling, so the entire pulse schedule can be encoded using at most 64 bits for inductive coupling, and 66 bits for capacitive coupling. 

In addition, the number of bits required to encode the ramps could be reduced by removing from the encoding bits the bits for which a kick is never applied in any of the optimized ramps (i.e., timings that correspond to a negative DRAG contribution). Furthermore, sharing pulse schedules between multiple qubits could result in a reduction in the encoding overhead for utility-scale quantum computers.

\section{Conclusion\label{sec:conclusion}}
In this work, we have presented a method to construct memory-efficient, high-fidelity, single-qubit gates for fluxonium qubits using SFQ control. We achieved average gate fidelities of 99.99\% in the case of inductive coupling and 99.9\% for capacitive coupling between the control chip and the qubits for a 128$\times$ frequency SFQ clock. We observe that leakage is the dominant source of coherent errors, and incoherent errors are of the same order of magnitude as those of coherent errors. The attained fidelities and the resulting error budget lead us to conclude that our approach is a promising and scalable alternative to traditional microwave control towards achieving fault-tolerant quantum computation. Future work on SFQ control of fluxoniums could include conducting stability analyses of the optimized pulse schedules or extending this approach to two-qubit gates.

\section*{ACKNOWLEDGEMENTS}

We thank our editor, Marko Bucyk, for his careful review and editing of the
manuscript. \mbox{M.~L.-M.} further acknowledges the financial support of Mitacs
and useful discussions with Baptiste Royer.
P.~R. acknowledges the financial support of Mike and Ophelia
Lazaridis, Innovation, Science and Economic Development Canada (ISED), and the
Perimeter Institute for Theoretical Physics. Research at the Perimeter Institute
is supported in part by the Government of Canada through ISED and by the
Province of Ontario through the Ministry of Colleges and Universities.



\clearpage

\bibliography{mabiblio}

\end{document}